\documentclass[preprint,5p,times,twocolumn]{elsarticle}

\usepackage{amsmath,amsfonts,amssymb}
\usepackage{algorithmic}
\usepackage{algorithm}
\usepackage{array}
\usepackage{subfig}
\usepackage{textcomp}
\usepackage{stfloats}
\usepackage{verbatim}
\usepackage{graphicx}
\usepackage{booktabs}

\usepackage[para]{threeparttable}
\usepackage{fontawesome5}
\usepackage{pifont}
\usepackage[table]{xcolor}
\usepackage{multirow,multicol}
\biboptions{square,sort&compress,comma,numbers}
\usepackage[acronym,toc]{glossaries}
\usepackage{tcolorbox}
\usepackage{xspace}
\usepackage{bm}
\usepackage{svg}
\usepackage[inline]{enumitem}
\usepackage{makecell}
\usepackage{todonotes}
\usepackage{flushend}

\newif\ifdraft
\draftfalse

\newacronym{genai}{GenAI}{Generative Artificial Intelligence}
\newacronym{traffgen}{NTG}{Network Traffic Generation}
\newacronym{traffclass}{NTC}{Network Traffic Classification}
\newacronym{id}{ID}{Intrusion Detection}
\newacronym{nda}{NDA}{Network Digital Assistance for Documentation \& Operation}
\newacronym{nid}{NID}{Network Intrusion Detection}
\newacronym{nids}{NIDS}{Network Intrusion Detection System}
\newacronym{ide}{IDE}{Implicit Density Estimation}
\newacronym{ede}{EDE}{Explicit Density Estimation}

\newacronym{la}{LA}{Log Analysis}
\newacronym{nla}{NSLA}{Networked System Log Analysis}
\newacronym{lad}{LAD}{Log-based Anomaly Detection}

\newacronym{xai}{XAI}{eXplainable Artificial Intelligence}
\newacronym{dl}{DL}{Deep Learning}
\newacronym{ai}{AI}{Artificial Intelligence}
\newacronym{iat}{IAT}{Inter Arrival Time}
\newacronym{tcpws}{TCP WS}{TCP Window Size}
\newacronym{pl}{PL}{Payload Length}
\newacronym{ps}{PS}{Packet Size}
\newacronym{ttl}{TTL}{Time To Live}
\newacronym{to}{TO}{Traffic Object}
\newacronym{tp}{TP}{Traffic Prediction}
\newacronym{ac}{AC}{Attack Classification}
\newacronym{ad}{AD}{Anomaly Detection}
\newacronym{nad}{NAD}{Network Anomaly Detection}
\newacronym{ntc}{NTC}{Network Traffic Classification}
\newacronym{nta}{NTA}{Network Traffic Analysis}
\newacronym{cnn}{CNN}{Convolutional Neural Network}
\newacronym{ml}{ML}{Machine Learning}
\newacronym{ids}{IDS}{Intrusion Detection System}
\newacronym{shap}{SHAP}{SHapley Additive exPlanations}
\newacronym{eda}{EDA}{Exploratory Data Analysis}
\newacronym{ae}{AE}{AutoEncoder}
\newacronym{ece}{ECE}{Expected Calibration Error}
\newacronym{ls}{LS}{Label Smoothing}
\newacronym{fl}{FL}{Focal Loss}
\newacronym{lime}{LIME}{Local Interpretable Model-agnostic Explanations}
\newacronym{qos}{QoS}{Quality of Service}
\newacronym{dt}{DT}{Decision Tree}
\newacronym{md}{MD}{Misuse Detection}
\newacronym{bmd}{bMD}{Binary Misuse Detection}
\newacronym{mmd}{mMD}{Multi-class Misuse Detection}
\newacronym{can}{CAN}{Controller Area Network}
\newacronym{apt}{APT}{Advanced Persistent Threat}

\newacronym{ran}{RAN}{Radio Access Network}
\newacronym{iot}{IoT}{Internet of Things}
\newacronym{llm}{LLM}{Large Language Model}
\newacronym{lstm}{LSTM}{Long Short-Term Memory}
\newacronym{xlstm}{xLSTM}{Extended Long Short-Term Memory}
\newacronym{tsne}{t-SNE}{t-distributed Stochastic Neighbor Embedding}
\newacronym{dir}{DIR}{Packet Direction}
\newacronym{darpa}{DARPA}{Defense Advanced Research Projects Agency}
\newacronym{nlp}{NLP}{Natural Language Processing}
\newacronym{sni}{SNI}{Server Name Indication}
\newacronym{svm}{SVM}{Support Vector Machine}
\newacronym{lrp}{LRP}{Layer-wise Relevance Propagation}
\newacronym{gradcam}{Grad-CAM}{Gradient-weighted Class Activation Mapping}
\newacronym{ig}{IG}{Integrated Gradients}
\newacronym{tls}{TLS}{Transport Layer Security}
\newacronym{dnn}{DNN}{Deep Neural Network}
\newacronym{som}{SOM}{Self Organizing Map}
\newacronym{knn}{KNN}{K-Nearest Neighbors}
\newacronym{pfi}{PFI}{Permutation Feature Importance}
\newacronym{ice}{ICE}{Individual Conditional Expectation}
\newacronym{pdp}{PDP}{Partial Dependence Plot}
\newacronym{cem}{CEM}{Contrastive Explanation Method}
\newacronym{coin}{COIN}{Contextual Outlier INterpretation}
\newacronym{ciu}{CIU}{Contextual Importance and Utility}
\newacronym{brcg}{BRCG}{Boolean Decision Rules via Column Generation}
\newacronym{cade}{CADE}{Contrastive Autoencoder for Drifting detection and Explanation}
\newacronym{deeplift}{DeepLIFT}{Deep Learning Important FeaTures}
\newacronym{lemna}{LEMNA}{Local Explanation Method using Nonlinear Approximation}
\newacronym{sdn}{SDN}{Software Defined Networking}
\newacronym{fpga}{FPGA}{Field Programmable Gate Array}
\newacronym{ddos}{DDos}{Distributed Denial of Service}
\newacronym{nmm}{NMM}{Network Monitoring and Management}
\newacronym{api}{API}{Application Programming Interface}
\newacronym{ui}{UI}{User Interface}
\newacronym{cli}{CLI}{Command Line Interface}
\newacronym{rnn}{RNN}{Recurrent Neural Network}
\newacronym{gru}{GRU}{Gated Recurrent Unit}
\newacronym{mlm}{MLM}{Masked Language Model}
\newacronym{dm}{DM}{Diffusion Model}

\newacronym{ddpm}{DDPM}{Denoising Diffusion Probabilistic Model}
\newacronym{ssm}{SSM}{State Space Model}
\newacronym{vit}{ViT}{Vision Transformer}

\newacronym{gan}{GAN}{Generative Adversarial Network}
\newacronym{vae}{VAE}{Variational Autoencoder}
\newacronym{gasf}{GASF}{Grammian Angular Summation Field}

\newacronym{jsd}{JSD}{Jensen-Shannon Divergence}
\newacronym{tvd}{TVD}{Total Variation Distance}
\newacronym{sr}{SR}{Success Rate}

\newacronym{rfc}{RFC}{Request for Comment}
\newacronym{cpsa}{CPSA}{Cryptographic Protocol Shapes Analyzer}
\newacronym{nf}{NF}{Network Function}
\newacronym{sml}{SLM}{Small Language Model}
\newacronym{gpp}{3GPP}{Third Generation Partnership Project}
\newacronym{rag}{RAG}{Retrieval-Augmented Generation}

\newacronym{fed}{FED}{Full Encoder-Decoder}
\newacronym{eo}{EO}{Encoder-Only}
\newacronym{do}{DO}{Decoder-Only}
\newacronym{sdp}{SDP}{Sequential Denoising Process}

\newacronym{mlp}{MLP}{MultiLayer Perceptron}

\newacronym{peft}{PEFT}{Parameter-Efficient Fine-Tuning}
\newacronym{ptq}{PTQ}{Post-Training Quantization}
\newacronym{lora}{LoRA}{Low-Rank Adaptation}
\newacronym{gguf}{GGUF}{GPT-Generated Unified Format}
\newacronym{nice}{NICE}{Non-linear Independent Components Estimation}
\newacronym{gpu}{GPU}{Graphics Processing Unit}
\newacronym{cvae}{CVAE}{Conditional Variational AutoEncoder}
\newacronym{gelu}{GELU}{Gaussian Error Linear Unit}
\newacronym{swiglu}{SwiGLU}{Swish-Gated Linear Units}
\newacronym{svd}{SVD}{Singular Value Decomposition}
\newacronym{rf}{RF}{Random Forest}
\newacronym{ood}{OOD}{Out-Of-Distribution}
\newacronym{pmf}{PMF}{Probability Mass Function}

\newacronym{rq}{RQ}{Research Question}
\newacronym{ra}{RA}{Research Answer}
\newacronym{ntg}{NTG}{Network Traffic Generation}

\usepackage{silence}
\usepackage{hyperref}
\usepackage{cleveref}

\usepackage{url}
\usepackage{comment}
\usepackage{dirtytalk}
\specialcomment{rewrite}{\begingroup\color{red}}{\endgroup}
\specialcomment{draft}{\begingroup\color{blue}}{\endgroup}
\specialcomment{passata1}{\begingroup\color{teal}}{\endgroup}
\specialcomment{anon}{\begingroup\color{orange}}{\endgroup}

\specialcomment{question}{\begingroup\color{black}\textit}{\endgroup}
\specialcomment{answer}{\begingroup\color{blue}}{\endgroup}
\specialcomment{future}{\begingroup\color{purple}}{\endgroup}
\excludecomment{future}

\specialcomment{draft}{\begingroup\color{blue}}{\endgroup}

\newcommand{\TCLIKE}%
{\mbox{small-increment}\xspace}



\usepackage{silence}
\usepackage{hyperref}
\usepackage{cleveref}
\usepackage{url}

\usepackage{todonotes}

\usepackage{comment}
\specialcomment{rewrite}{\begingroup\color{red}}{\endgroup}
\specialcomment{draft}{\begingroup\color{blue}}{\endgroup}
\specialcomment{passata1}{\begingroup\color{teal}}{\endgroup}

\specialcomment{question}{\begingroup\color{black}\textit}{\endgroup}
\specialcomment{answer}{\begingroup\color{blue}}{\endgroup}
\specialcomment{future}{\begingroup\color{purple}}{\endgroup}
\excludecomment{future}

\newcommand{\nota}[1]{\color{red}{#1}\xspace}
\renewcommand{\nota}[1]{}

\newcommand{\MIRAGE}{\mbox{$\mathtt{Mirage}$-$\mathtt{2019}$}\xspace}
\newcommand{\CESTLS}{\mbox{$\mathtt{CESNET}$-$\mathtt{TLS22}$}\xspace}
\newcommand{\TLSETY}{\mbox{$\mathtt{CESNET}$-$\mathtt{TLS22}$-$\mathtt{80}$}\xspace} 
\newcommand{\eg}{e.g.,\xspace}
\newcommand{\ie}{i.e.\xspace}

\newcommand{\RF}{\texttt{RF}\xspace}

\newcommand{\CNN}{\texttt{2D-CNN}\xspace}

\newcommand{\CVAE}{\fmtTT{CVAE}\xspace}
\newcommand{\NETDIFFUS}{\fmtTT{NetDiffus}\xspace}
\newcommand{\NETDIFFUSNR}{\fmtTT{NetDiffus++}\xspace}
\newcommand{\NETDIFFUSION}{\fmtTT{NetDiffusion}\xspace}
\newcommand{\MAMBA}{\fmtTT{Mamba}\xspace}

\newcommand{\LLAMA}{\fmtTT{LLaMA}\xspace}

\newcommand{\PLDIR}{$\mathtt{PL}\times\mathtt{DIR}$\xspace}

\newtcolorbox{recap}[1][]{%
  enhanced jigsaw,
  sharp corners,
  colback=white,
  borderline={1pt}{-2pt}{black},
  fontupper={\setlength{\parindent}{20pt}},
  #1
}

\newcommand{\processWord}[1]{%
    \begingroup
    \def\processChar##1{%
        \ifx##1\relax
        \else
            \ifx##1-%
                \text{-}%
            \else
                \ifx##1é
                    \mathtt{\acute{e}}%
                \else
                    \ifx##1\'a
                        \acute{a}%
                    \else
                        \ifx##1\'i
                            \acute{i}%
                        \else
                            \ifx##1\'o
                                \acute{o}%
                            \else
                                \ifx##1\'u
                                    \acute{u}%
                                \else
                                    \ifx##1\`e
                                        \grave{e}%
                                    \else
                                        \ifx##1\`a
                                            \grave{a}%
                                        \else
                                            \ifx##1\`i
                                                \grave{i}%
                                            \else
                                                \ifx##1\`o
                                                    \grave{o}%
                                                \else
                                                    \ifx##1\`u
                                                        \grave{u}%
                                                    \else
                                                        \mathtt{##1}%
                                                    \fi
                                                \fi
                                            \fi
                                        \fi
                                    \fi
                                \fi
                            \fi
                        \fi
                    \fi
                \fi
            \fi
            \expandafter\processChar
        \fi
    }%
    \expandafter\processChar#1\relax
    \endgroup
}

\newcommand{\fmtTT}[1]{%
    \begingroup
    \def\split##1 ##2\relax{%
        \processWord{##1}%
        \ifx##2\empty%
        \else
            \:%
            \expandafter\split\expandafter##2\relax
        \fi
    }%
    $\split#1 \relax$%
    \endgroup
}

\newcommand{\subsec}[1]{%
    \vspace{5pt}%
    \noindent%
    \textbf{#1}%
}

\journal{Computer Networks}

\begin{document}

\title{
Lightweight Generative AI for Network Traffic Generation:\\
Fidelity, Augmentation, and Classification
}

\author[pegaso]{Giampaolo Bovenzi\fnref{fn:bovenzi}}
\ead{giampaolo.bovenzi@unipegaso.it}

\author[unina]{Domenico Ciuonzo}
\ead{domenico.ciuonzo@unina.it}

\author[huawei]{Jonatan Krolikowski}
\ead{jonatan.krolikowski@huawei.com}

\author[unina]{Antonio Montieri\corref{cor1}}
\ead{antonio.montieri@unina.it}

\makeatletter
\g@addto@macro\elsauthors{\unskip,\par\let\authorsep\@empty}
\makeatother

\author[unina]{Alfredo Nascita}
\ead{alfredo.nascita@unina.it}

\author[unina]{Antonio Pescap{\`e}}
\ead{pescape@unina.it}

\author[huawei]{Dario Rossi}
\ead{dario.rossi@huawei.com}

\address[pegaso]{Department of Information Science and Technology, Pegaso Telematic University, Italy}
\address[unina]{Department of Electrical Engineering and Information Technologies, University of Naples Federico II, Italy}
\address[huawei]{Huawei Technologies France SASU, France}

\fntext[fn:bovenzi]{The work reported in this paper was conducted while Giampaolo Bovenzi was affiliated with the Department of Electrical Engineering and Information Technologies, University of Naples Federico II, Italy.}
\cortext[cor1]{Corresponding author.}

\begin{abstract}
\gls{ntc} increasingly relies on data-driven models, yet its practical deployment is often constrained by limited labeled data, strict privacy requirements, and the cost of collecting representative traffic traces.
While \gls{ntg} provides an effective means to mitigate data scarcity, conventional generative methods struggle to model the complex temporal dynamics of modern traffic and often incur high computational costs.
In this article, we investigate lightweight \gls{genai} architectures for practical \gls{ntg}. 
Rather than generating raw packet bytes or relying on large foundation models, we synthesize compact flow-level traffic representations derived from early packet-header information, enabling transformer-based, state-space, and diffusion models with only a few million parameters. 
We present a modular \gls{genai} pipeline for \gls{ntg} and evaluate it along four complementary axes: 
($i$) synthetic traffic fidelity, 
($ii$) synthetic-only training for reduced raw-data sharing in \gls{ntc},
($iii$) data augmentation under low-data regimes, 
and ($iv$) computational efficiency.
Experiments on two heterogeneous datasets show that lightweight transformer-based and state-space models preserve both static and temporal traffic characteristics, while providing useful synthetic data for downstream \gls{ntc}.
Among them, transformer-based models offer the best fidelity--efficiency trade-off, combining high-quality traffic generation
with moderate computational overhead.

\end{abstract}


\maketitle

\glsresetall

\section{Introduction}
\label{sec:introduction}
\begin{figure*}[htp]
    \centering

    \includegraphics[trim={20 20 20 20},clip,width=\linewidth]{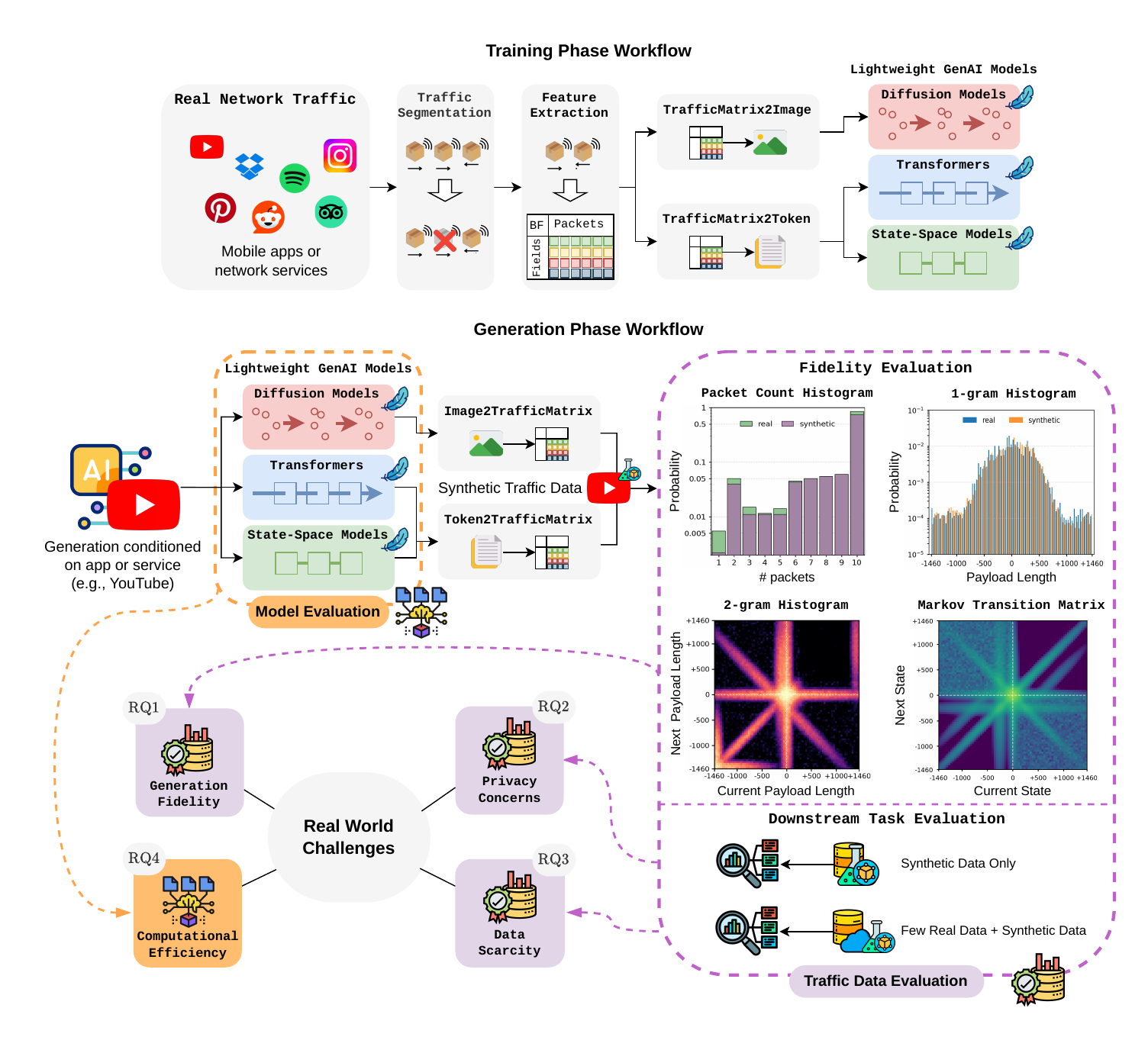}
    
    \caption{
    Overview of the proposed \textbf{lightweight GenAI pipeline for network traffic generation} and \textbf{real-world challenges} addressed by our \textbf{Research Questions}.
    \emph{Training phase workflow:} real network traces are segmented into biflows and mapped into canonical image- or token-based representations; these serve as inputs to train diffusion, transformer, or state-space generative models.
    \emph{Generation phase workflow:} trained GenAI models are conditioned to generate image- or token-based representations for a target traffic class; these are then converted back into traffic matrices.
    Generation efficacy is assessed in terms of both \emph{traffic data evaluation} (in \textcolor{violet}{\textbf{violet}}) and \emph{model evaluation} (in \textcolor{orange}{\textbf{orange}}). 
    Traffic data evaluation comprises fidelity and downstream task assessment, while model evaluation focuses on computational efficiency and resource usage.
    }
    \label{fig:ntg_pipeline}
\end{figure*}

The rapid evolution of \gls{genai} is reshaping communications, accelerating the vision of autonomous and self-evolving networks~\cite{bariah2024large}.
Large generative models have demonstrated remarkable potential across telecom domains, exploiting massive datasets to learn complex patterns and generate new content.
Applications already span physical-layer optimization and channel modeling~\cite{xu2025generative}, and are increasingly extending to \gls{genai}-enabled network operations and optimization for future 6G networks~\cite{long20246g}.
While adoption for data-centric network monitoring and management is also blooming~\cite{bovenzi2025mapping}, it remains an open frontier~\cite{perez2026network}.

Within this landscape, \gls{ntc} remains pivotal for network management and security across mobile, wired, and capillary \gls{iot} environments. 
While accurate \gls{ntc} enables various critical functions, such as anomaly detection, quality of service provisioning, and security policy enforcement, traditional data-driven solutions often struggle to meet practical desiderata. 
Despite the advancements fueled by \gls{ml}, practical constraints like limited labeled data, imbalanced classes, and strict privacy regulations act as severe roadblocks.
These constraints hinder scalability and reduce robustness in real-world settings~\cite{aceto2023ai}.

To address these limitations, \gls{ntg} has been widely explored to synthesize realistic traffic data, enrich training sets, and strengthen classification robustness. 
Nevertheless, conventional approaches fall short. 
Classical data augmentation and traditional \gls{ml}-based \gls{ntc} rely on handcrafted statistical features, limiting adaptability to evolving traffic patterns and failing to capture rich temporal dependencies. 
Meanwhile, conventional \gls{ntg} solutions struggle with a core trade-off: producing high-fidelity synthetic traffic while maintaining the computational efficiency required by modern networks~\cite{bovenzi2025mapping}.

In this article, we advocate for a shift toward \textbf{lightweight \gls{genai} architectures} for practical \gls{ntg}.
As depicted in Fig.~\ref{fig:ntg_pipeline}, we propose a modular \gls{genai} pipeline specifically tailored for \gls{ntg}.
Rather than synthesizing fine-grained payload bytes using massive foundation models, our approach generates \emph{a compact traffic representation derived from header fields of the first packets in each network flow}.
This choice enables training and operating lightweight configurations of modern \gls{genai} models, namely \emph{Transformers}, \emph{\glspl{ssm}}, and \emph{\glspl{dm}}, under a budget of \emph{$1$--$2$ million parameters}, ensuring computational efficiency without sacrificing generation fidelity.

To validate this lightweight paradigm, we formulate \textbf{four \glspl{rq}} aligned with real-world networking challenges (Fig.~\ref{fig:ntg_pipeline}, bottom left): 
\begin{enumerate}[leftmargin=25pt,label=\emph{\colorbox{gray!25}{\gls{rq}\arabic*:}}]
\item Can lightweight \gls{genai} faithfully reproduce real traffic patterns, and how should generation fidelity be evaluated?
\item Can \gls{genai} synthetic traffic support reduced raw-data sharing for downstream \gls{ntc} while retaining classification utility?
\item Can \gls{genai} synthetic traffic mitigate training-data scarcity for downstream \gls{ntc} in low-data regimes?
\item Can lightweight \gls{genai} models support deployment-oriented \gls{ntg} with practical computational costs?
\end{enumerate}
In other words, we assess both generated data quality from different viewpoints (\emph{Traffic Evaluation} -- \gls{rq}1 to \gls{rq}3) and \gls{genai} model efficiency (\emph{Model Evaluation} -- \gls{rq}4).

To answer these \glspl{rq}, we make the following \textbf{contributions}: 
\begin{itemize}
    \item we \emph{design a modular lightweight \gls{genai} pipeline} for \gls{ntg} and define a \emph{fidelity assessment methodology} to evaluate generated traffic realism and quality; 
    \item we investigate the effectiveness of \emph{synthetic-only training} for reduced raw-data sharing in \gls{ntc}; 
    \item we evaluate the impact of \emph{synthetic traffic augmentation} on \gls{ntc} in low-data regimes; 
    \item we assess the \emph{deployment feasibility} of \gls{genai}-based traffic generators via computational efficiency and resource usage.
\end{itemize}

Experiments on two public datasets---\MIRAGE ($40$ mobile apps) and \TLSETY ($80$ network services)---demonstrate that lightweight \gls{genai} models achieve strong fidelity, preserving both static and temporal traffic patterns.
Compared with traditional \gls{ntg} baselines (\eg \CVAE, SMOTE, and domain-expert transformations) and \glspl{dm}, Transformer- and \gls{ssm}-based approaches (\ie \LLAMA and \MAMBA, respectively) exhibit higher performance. 
Traffic classifiers trained exclusively on synthetic data reach up to $87\%$ F1-score on real traffic, with an absolute gap below $5\%$
compared with real-data training, while \gls{genai}-driven augmentation improves F1-score values by up to $40\%$ in low-data regimes.
Last but not least, resource analysis reveals a clear trade-off between architectural complexity and computational cost, with Transformer-based models offering the most favorable balance for practical deployment, combining a moderate memory footprint with the lowest generation latency among the high-fidelity generators.

The remainder of the article explores the background and related work shaping the current \gls{ntg} landscape (Sec.~\ref{sec:back_rw}) and presents the proposed lightweight \gls{genai} pipeline (Sec.~\ref{sec:methodology}). 
It then details the experimental setup (Sec.~\ref{sec:experimental_setup}) and discusses the results, addressing the four \glspl{rq} (Sec.~\ref{sec:experimental_results}), before concluding with directions for future work (Sec.~\ref{sec:conclusion}).

\section{The \gls{genai} Paradigm Shift in Network Traffic Generation}
\label{sec:back_rw}
This section frames \gls{ntg} evolution, categorizing existing work into three phases: early statistical methods and conventional \gls{ml} models (Sec.~\ref{subsec:rw_traditional}), current large-scale \gls{genai} models (Sec.~\ref{subsec:rw_recent}), and the emerging need for lightweight solutions addressing our research gap (Sec.~\ref{subsec:positioning}).

\subsection{Traditional Generative Methods}
\label{subsec:rw_traditional}
Early \gls{ntg} relied on \emph{statistical generative models} (\eg Markov models~\cite{aceto2021characterization}) to capture sequential dependencies through state transitions. 
Though intuitive and lightweight, they do not scale well to mid- or long-range dependencies.
\emph{Conventional \gls{ml} models} improved expressive power~\cite{aceto2024synthetic}.
Variational Autoencoders learn probabilistic latent representations that enhance reconstruction accuracy~\cite{meslet2022necstgen}, while Normalizing Flows provide exact likelihood estimation at the cost of a higher computational overhead~\cite{gudovskiy2022cflow}.
Generative Adversarial Networks, in contrast, focus on high-fidelity sampling~\cite{madarasingha2022videotrain} but often suffer from training instability and mode collapse~\cite{hui2022knowledge}. Although traditional methods support downstream tasks like intrusion detection and traffic classification~\cite{bovenzi2025mapping, aceto2024synthetic}, they struggle to balance training stability, computational efficiency, and fidelity in modeling complex, evolving traffic patterns.

\subsection{Recent GenAI Advancements}
\label{subsec:rw_recent}
Since 2021, \gls{ntg} has been impacted by the growing popularity of \emph{Transformers}, \emph{\glspl{dm}}, and \emph{\glspl{ssm}}, each employing distinct strategies for traffic representation~\cite{perez2026network}.
Originating from \gls{nlp}, Transformer-based models (\eg \fmtTT{GPTs} and \fmtTT{T5})~\cite{bikmukhamedov2021, Kholgh2023PACGPT, qu2024trafficgpt, cui2025trafficllm, mayhoub2026talk, delgado2025gpt} and \glspl{ssm} (\ie \fmtTT{Mamba})~\cite{chu2024feasibility} treat network traffic as token sequences, learning networking ``grammar'' to model flow dynamics.
Input sequences range from raw packet bytes~\cite{qu2024trafficgpt, chu2024feasibility} and header fields (\eg packet sizes, inter-arrival times, and directions)~\cite{bikmukhamedov2021} to \texttt{tcpdump}/\texttt{tshark} packet summaries~\cite{Kholgh2023PACGPT,mayhoub2026talk,delgado2025gpt}, while some approaches also generate executable \texttt{Scapy} code to construct PCAP traces~\cite{Kholgh2023PACGPT,delgado2025gpt}.
To further bridge the modality gap between natural language and network data, the authors of~\cite{cui2025trafficllm} combine specialized traffic-domain tokenization and multimodal learning to understand expert instructions and learn task-specific traffic representations simultaneously.
Crucially, generation strategies operate either iteratively, constructing flows ``token-by-token'' akin to sentences in \gls{nlp}~\cite{qu2024trafficgpt}, or natively synthesize complete sequences~\cite{bikmukhamedov2021, chu2024feasibility}. 
A distinct approach is adopted in~\cite{Kholgh2023PACGPT,delgado2025gpt}, where \gls{genai} models generate executable Python code using the \texttt{Scapy} library to construct network packets and traces, rather than directly synthesizing traffic representations.

Conversely, \glspl{dm}, originating from computer vision, have recently gained traction across the entire networking stack, from physical-layer channel generation and resource management~\cite{xu2025generative} to \gls{ntg}~\cite{sivaroopan2023netdiffus, jiang2024netdiffusion}. 
\glspl{dm} synthesize traffic by iteratively denoising random data until structured patterns emerge. 
This requires encoding traffic into image-like representations~\cite{sivaroopan2023netdiffus} or directly modeling raw byte streams~\cite{jiang2024netdiffusion} to capture high-fidelity details.
Although these ``large'' models achieve high fidelity, their byte-level processing and iterative generation often make them impractical for network deployment.

\subsection{Positioning of Our Work}
\label{subsec:positioning}
Our work diverges from the emerging trend of massive network foundation models~\cite{mayhoub2026talk} to investigate the feasibility of \emph{lightweight \gls{genai}}. 
Indeed, while foundation models offer generalization and fine-tuning capabilities, they impose prohibitive computational costs.
Instead, we explore \gls{ntg} solutions based on Transformers, \glspl{ssm}, and \glspl{dm} under strict resource constraints (\ie~$1$--$2$M parameters), prioritizing deployability and training/inference speed.

Unlike related works focused on generating raw packet bytes~\cite{jiang2024netdiffusion, chu2024feasibility} (which increases complexity and risks payload data leakage), we synthesize lightweight traffic features constituting the network fingerprint of applications. 
Specifically, we model the time series of payload lengths and packet directions (represented as \PLDIR), akin to~\cite{qu2024trafficgpt, bikmukhamedov2021, sivaroopan2023netdiffus}, but explicitly discard inter-arrival times, as these depend on network conditions rather than application logic~\cite{wang2024data}. 

Furthermore, we address the lack of rigorous validation in prior studies by introducing advanced fidelity metrics, including \PLDIR $n$-grams and Markov transition matrices, to assess temporal integrity. 

Finally, targeting practical downstream tasks for \gls{ntg}, our comprehensive evaluation demonstrates that lightweight \gls{genai} enables effective classifier training even in ($a$) synthetic-only or ($b$) low-data regimes, ensuring efficiency for real-world deployment.

\section{Lightweight \gls{genai}-based \gls{ntg} Pipeline at Work}
\label{sec:methodology}
Figure~\ref{fig:ntg_pipeline} depicts the modular \gls{ntg} pipeline powered by lightweight \gls{genai} models, structured around \emph{two distinct phases}. 
The \emph{training phase} (Sec.~\ref{subsec:training_phase}) pre-processes real network traces into suitable representations for training \gls{genai} models, while the \emph{generation phase} (Sec.~\ref{subsec:generation_phase}) leverages them to synthesize high-fidelity, application-conditioned traffic data. 
Finally, Sec.~\ref{subsec:genai_models} introduces the specific lightweight \gls{genai} architectures used to implement the pipeline.

\subsection{Training Phase Workflow}
\label{subsec:training_phase}
The training phase ingests \emph{real network traffic traces} (\eg from mobile apps or network services). 
To bypass the computational overhead and privacy risks associated with raw payload utilization, data undergoes \emph{traffic segmentation} to group packets into bidirectional flows (biflows)%
\footnote{A biflow is a network flow consisting of packets flowing bidirectionally between the same network and transport endpoints---IPs, ports, and L4 protocol. It represents both directions of communication as a single entity.}, 
followed by \emph{feature extraction}. 
This produces a highly efficient \emph{traffic matrix} representation, where rows correspond to packets and columns encode packet fields. 
Specifically, we extract the \gls{pl} and \gls{dir} from the initial packets of each biflow.

Depending on the \gls{genai} model employed, this matrix undergoes a specific modality-mapping before training:
\begin{itemize}
    \item \textbf{\texttt{TrafficMatrix2Image}:} Image-based models, such as \glspl{dm}, interpret the traffic matrix, possibly after model-specific preprocessing, as a structured 2D image. This enables \glspl{dm} to learn complex 2D patterns that encode the underlying traffic features and sequential dependencies across packets, generating the entire traffic representation as a whole rather than autoregressively token by token.
    \item \textbf{\texttt{TrafficMatrix2Token}:} Sequence-based models, such as Transformers and \glspl{ssm}, represent the traffic matrix as a sequence of discrete tokens, where each token jointly encodes the packet-level fields at one sequence step. This formulation supports autoregressive generation and effectively captures packet-to-packet dependencies underlying network traffic dynamics.
\end{itemize}
Given these mapped representations, the goal of the selected \gls{genai} models is to learn a distribution that faithfully captures the underlying structure of the real network traces.

\subsection{Generation Phase Workflow}
\label{subsec:generation_phase}
Once the \gls{genai} models are trained, they are employed to synthesize new traffic samples. 
Generation is \emph{conditioned} on a \texttt{<CLASS>} identifier specifying the target network class (\eg a certain mobile app like YouTube, or a network service).
The \gls{genai} architectures generate synthetic samples in their native formats: 2D images for \glspl{dm} or token sequences for Transformers and \glspl{ssm}. 
To leverage these outputs for downstream \gls{ntc}, an inverse mapping step is needed. 
Specifically, the \textbf{\texttt{Image2TrafficMatrix}} and \textbf{\texttt{Token2TrafficMatrix}} steps reconstruct the generated samples back into the original \emph{traffic matrix} format, recovering the corresponding synthetic \glspl{pl} and \glspl{dir}.

Generation efficacy is evaluated from two complementary viewpoints:
\begin{itemize}
    \item \textbf{Traffic Evaluation:} To assess generation fidelity, we quantify the divergence between real and synthetic traffic distributions using measures such as the \gls{jsd}. More precisely, we compute these divergences across the traffic properties illustrated in Fig.~\ref{fig:ntg_pipeline}: \emph{packet count histograms} for biflow-level behavior, \emph{1-gram histograms} for marginal distributions of \gls{pl} and \gls{dir}, \emph{2-gram histograms} for sequential dependencies across consecutive packet pairs, and \emph{Markov transition matrices} for first-order transition dynamics. 
    Beyond fidelity, we evaluate the utility of generated traffic in two practical downstream \gls{ntc} scenarios: \emph{synthetic-only training}, where classifiers are trained exclusively on synthetic data and tested on real samples, and \emph{data augmentation}, assessing whether enriching a few real samples with synthetic ones boosts classification performance.
    
    \item \textbf{Model Evaluation:} To assess the deployment feasibility of our lightweight \gls{genai} models, we also profile their \emph{computational efficiency} during training and inference, evaluating training time, generation latency, GPU/CPU utilization, memory consumption, and on-disk model size. 
    Additionally, we investigate post-training quantization to assess whether the most promising generator can be further optimized for resource-constrained environments.
\end{itemize}

\begin{figure*}[t]
    \centering
    \subfloat[\MIRAGE\label{fig:mirage}]{\includegraphics[height=.31\columnwidth]{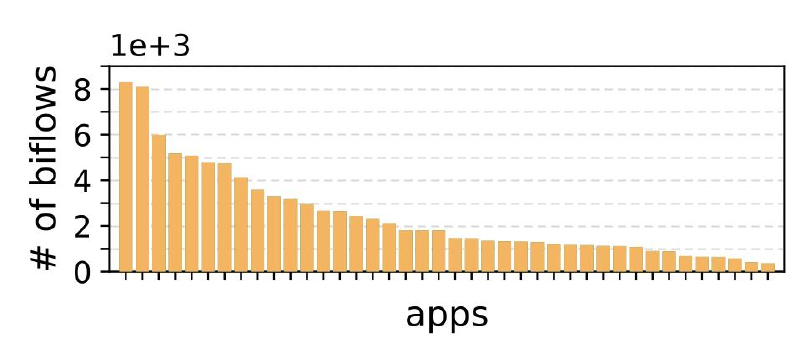}}
    \subfloat[\TLSETY\label{fig:cesnet}]{\includegraphics[height=.31\columnwidth, trim={30 0 0 0}, clip]{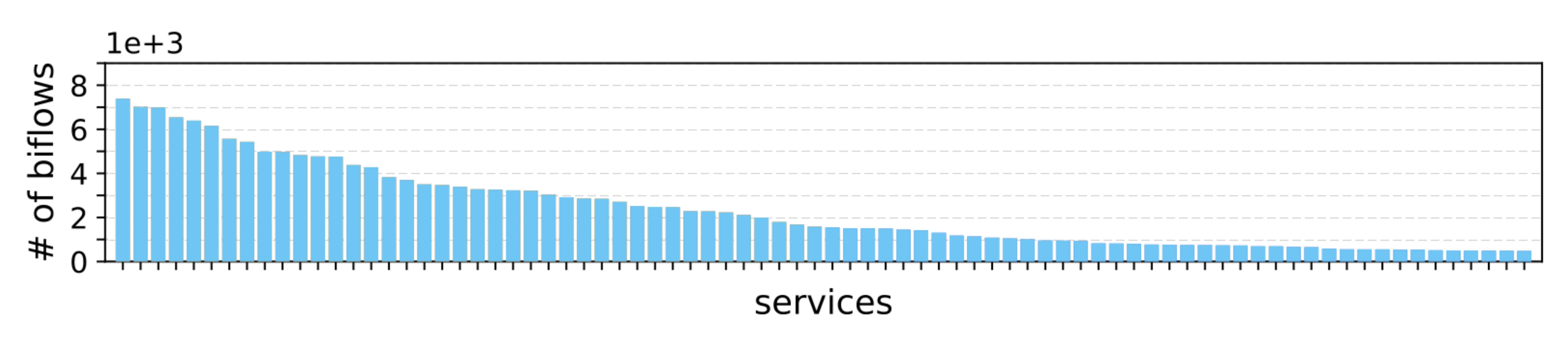}}
    \caption{Datasets composition in terms of the number of biflows for each app or service.}
    \label{fig:dataset}
\end{figure*}

\subsection{Lightweight GenAI Models}
\label{subsec:genai_models}
To implement our \gls{ntg} pipeline under strict computational constraints, we leverage \emph{natively lightweight \gls{genai} models} ($\approx\!1\text{--}2$M parameters by design) belonging to different families:
\begin{itemize}
    \item \textbf{Diffusion Models (DMs):} \glspl{dm} generate synthetic samples by iteratively reversing a noising process. We employ \NETDIFFUSNR, an enhanced version of \NETDIFFUS~\cite{sivaroopan2023netdiffus} that applies diffusion to a \gls{gasf}-based transformation of the traffic matrix. Unlike the original approach, \NETDIFFUSNR incorporates a post-generation refinement stage that maps generated \gls{gasf} images back to traffic sequences while minimizing reconstruction error, with the goal of improving generation fidelity.%
    \footnote{We exclude heavier byte-level alternatives like \NETDIFFUSION~\cite{jiang2024netdiffusion} from our evaluation, as their scale (hundreds of millions of parameters) and simplified direction modeling are inconsistent with our lightweight, flow-level, time-series-oriented design.}
    
    \item \textbf{Transformer-based Models:} Transformers excel at sequence modeling by capturing dependencies across tokens through self-attention. We adopt a lightweight configuration of \LLAMA~\cite{touvron2023llama}, a causal decoder-only architecture designed to capture long-range dependencies across sequence positions. It leverages causal self-attention to autoregressively synthesize packet-level traffic sequences.
    
    \item \textbf{Structured State-Space Models (SSMs):} To complement Transformers, we explore \glspl{ssm} designed for efficient sequence processing. We employ \MAMBA~\cite{chu2024feasibility}, which replaces standard attention with a selective state-space formulation. Operating causally, it achieves linear-time sequence processing, making it well-suited for modeling per-biflow traffic sequences. 
\end{itemize}

\section{Experimental Setup}
\label{sec:experimental_setup}
This section describes the experimental setup adopted to evaluate our lightweight \gls{genai}-based \gls{ntg} pipeline. 
We first introduce the two real-world network traffic datasets and their characteristics (Sec.~\ref{subsec:datasets}). 
We then describe the traffic representation, preprocessing steps, and generative model configurations used in our experiments (Sec.~\ref{subsec:configuration}).

\subsection{Datasets}
\label{subsec:datasets}
Our evaluation relies on two public network traffic datasets:
\MIRAGE%
\footnote{\url{https://traffic.comics.unina.it/mirage/mirage-2019.html}}
and \TLSETY, derived from \CESTLS%
\footnote{\url{https://www.liberouter.org/datasets/cesnet-tls22}}.

\subsec{\MIRAGE}~\cite{aceto2019mirage}.
This dataset contains traffic collected from $40$ Android apps of $16$ different categories (according to the Google Play Store).
It was collected at the ARCLAB laboratory of the University of Napoli Federico II between May 2017 and May 2019. 
\MIRAGE comprises over $4,600$ PCAP traces, each captured during a $5$--$10$ minute session, and approximately $100$\,k labeled biflows.
Although the number of capture sessions is comparable across applications, the biflow distribution is heavy-tailed: 
the $10$ and $20$ most represented applications account for $54.2\%$ and $78.9\%$ of all biflows, respectively.
The class imbalance ratio, defined as the ratio between the number of biflows in the largest and smallest classes, is $22.98$.
Figure~\ref{fig:mirage} shows the class-wise biflow distribution.

\subsec{\TLSETY}~\cite{luxemburk2023fine}.
The original \CESTLS dataset includes traffic data generated by $191$ services classified into $21$ distinct categories. 
It was collected during the first two weeks of October 2021 using high-speed probes deployed at the perimeter of the \emph{CESNET2} network, monitoring multiple $100$\,Gb/s links.
The original dataset comprises approximately $140$\,M biflows labeled using the \gls{sni} field from \gls{tls} handshakes. 
For our evaluation, we construct \TLSETY by selecting the $80$ most represented services from \CESTLS and randomly retaining $0.15\%$ of the biflows in each selected class, obtaining a number of samples comparable to \MIRAGE.
The resulting class imbalance ratio is $15.40$, and the corresponding biflow distribution is shown in Fig.~\ref{fig:cesnet}.
Together, \MIRAGE and \TLSETY provide evaluation scenarios with $40$ and $80$ traffic classes, respectively, enabling us to assess \gls{ntg}
in heterogeneous multi-class settings that are broader than prior \gls{genai}-based \gls{ntg} benchmarks focusing on fewer applications,
services, or attacks~\cite{sivaroopan2023netdiffus,chu2024feasibility,mayhoub2026talk}.

For both datasets, we adopt a stratified biflow-level $70\%/30\%$ train/test split, preserving the class distribution across the two subsets.
The same split is used for all considered models to ensure a consistent experimental comparison.

\newcommand{\faHuggingFace}{{\includesvg[height=1em]{images/icons/hf-logo.svg}}}
\newcommand{\orange}{\cellcolor[HTML]{FAC487}}
\newcommand{\blue}{\cellcolor[HTML]{87D4FA}}

\begin{table}[t]
\centering
\caption{
Configuration of the considered lightweight GenAI models, including architectural hyperparameters and trainable parameter counts for \MIRAGE{} and \TLSETY{}.
}
\label{tab:genai_models_config}
\setlength{\tabcolsep}{3pt}
\resizebox{\columnwidth}{!}{
\begin{threeparttable}
\begin{tabular}{lccccccc}
\toprule
\textbf{Model} &
\textbf{HS} &
\textbf{IS} &
\textbf{\#L} &
\textbf{\#AH} &
\orange\textbf{\MIRAGE} &
\blue\textbf{\TLSETY} &
\textbf{Repo}\\
\midrule

\CVAE
& 500/250
& 20
& 6
& --
& \orange 1.01M + 0
& \blue 1.09M + 0
& -- \\

\NETDIFFUSNR
& 32
& --
& 4
& 4
& \orange 1.02M + 5k
& \blue 1.03M + 10k
& \href{https://github.com/Nirhoshan/NetDiffus}{\faGithub} \\

\MAMBA
& 72
& 144
& 4
& --
& \orange 966k + 427k
& \blue 966k + 432k
& \href{https://huggingface.co/docs/transformers/main/model_doc/mamba}{\faHuggingFace} \\

\LLAMA
& 160
& 320
& 4
& 8
& \orange 1.03M + 948k
& \blue 1.03M + 961k
& \href{https://huggingface.co/docs/transformers/main/model_doc/llama}{\faHuggingFace} \\

\bottomrule
\end{tabular}

\begin{tablenotes}[flushleft]
\textbf{Legend} --
\textbf{HS}: Hidden Size;
\textbf{IS}: Intermediate Size;
\textbf{\#L}: Number of Layers;
\textbf{\#AH}: Number of Attention Heads;
\MIRAGE/\TLSETY{} columns report the number of trainable parameters as the sum of \emph{non-embedding + embedding} parameters;
\textbf{Repo}: Repository (clickable icon).

\end{tablenotes}
\end{threeparttable}
}

\end{table}

\subsection{Traffic Representation and Model Configuration}
\label{subsec:configuration}
Raw traffic data are pre-processed into sequences of signed \glspl{pl} ($\pm$\glspl{pl}), where negative and positive values encode downstream and upstream \glspl{dir}, respectively.
We focus on the first $10$ non-zero-payload packets of each biflow to capture the initial communication dynamics, including patterns associated with protocol establishment (\eg TLS and QUIC) and early application exchanges, while keeping the representation compact. 
We consider only payload \glspl{pl} and \glspl{dir}, which characterize the size and directionality of packet exchanges. 
The exclusion of \glspl{iat} is further motivated by prior data-driven explainability studies showing that timing features generally contribute less to traffic classification than packet-size and direction-related features~\cite{nascita2021xai}.

For a fair cross-architecture comparison, all considered \gls{genai} models are \emph{natively lightweight} and constrained to a common budget of $1$--$2$M trainable parameters. 
Alongside these models, we employ a Conditional Variational Autoencoder (\CVAE) baseline for class-aware traffic generation.
Table~\ref{tab:genai_models_config} summarizes their key hyperparameters.
For sequence-based models, we construct a vocabulary by assigning a unique token to each possible signed integer-valued \gls{pl}. 
This yields $2 \times \mathrm{PL}_{\max}$ tokens before adding special tokens, with $\mathrm{PL}_{\max}$ denoting the maximum considered payload length.
This direct integer mapping requires no additional quantization, thus avoiding precision loss. 
The vocabulary is augmented with an \texttt{<EOS>} token, a \texttt{<PAD>} token, and $\mathrm{N}$ \texttt{<CLASS>} tokens ($N=40$ for \MIRAGE and $N=80$ for \TLSETY). 
To ensure fixed-length training inputs, biflows with fewer than $10$ non-zero-payload packets are right-padded with \texttt{<PAD>} tokens.

\section{Experimental Results}
\label{sec:experimental_results}
This section presents and discusses the experimental results, providing \emph{Research Answers} (RAs) to the four Research Questions introduced in Sec.~\ref{sec:introduction}.
We first assess synthetic traffic fidelity (RA1, Sec.~\ref{subsec:ra1_fidelity}), then examine the utility of generated traffic for downstream \gls{ntc} through synthetic-only training (RA2, Sec.~\ref{subsec:ra2_tstr}) and data augmentation under limited real-data availability (RA3, Sec.~\ref{subsec:ra3_augmentation}).
Finally, we evaluate the computational efficiency and deployment feasibility of lightweight \gls{genai} models (RA4, Sec.~\ref{subsec:ra4_computational}).

\begin{figure}[t]
    \centering
    \includegraphics[width=0.95\columnwidth,trim=0 40 0 0,clip]{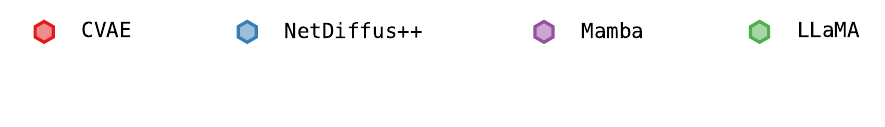}
    
    \subfloat[\MIRAGE]{
    \includegraphics[width=0.48\columnwidth,trim=5 7 7 7,clip]{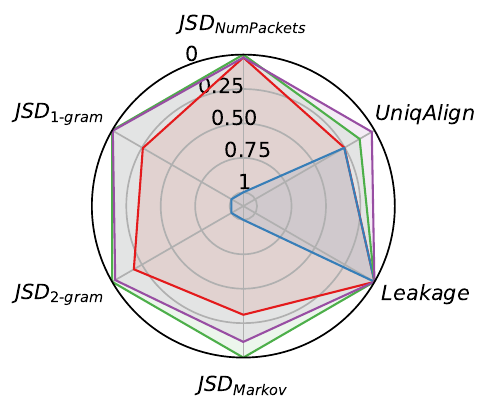}
    }
    \subfloat[\TLSETY]{
    \includegraphics[width=0.48\columnwidth,trim=5 7 7 7,clip]{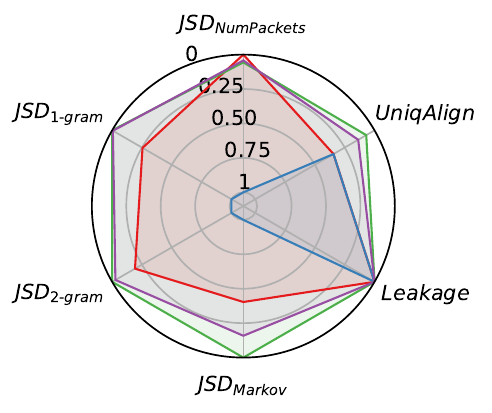}
    }  
    \caption{Radar plots of $6$ fidelity metrics comparing real and synthetic traffic data across generative models for \MIRAGE (left) and \TLSETY (right). For all considered metrics, lower values indicate better performance (with $0$ being optimal). Note that the axes are scaled with $0$ at the outer edge, meaning that models producing larger polygon areas exhibit more favorable overall generation profiles.
    }
    \label{fig:radar_fidelity}
\end{figure}

\subsection{RA1 -- Fidelity Evaluation} 
\label{subsec:ra1_fidelity}
To answer \gls{rq}1, we quantitatively assess the fidelity of generated traffic through the six metrics reported in the radar plots of Fig.~\ref{fig:radar_fidelity}. 
First, we translate the traffic properties illustrated in Fig.~\ref{fig:ntg_pipeline} into numerical metrics by computing the macro-averaged \gls{jsd} between real and synthetic distributions for: ($a$) packet count histograms ($\text{JSD}_{\text{NumPackets}}$), ($b$) 1-gram histograms ($\text{JSD}_{\text{1-gram}}$) capturing individual signed \gls{pl} values, ($c$) 2-gram histograms ($\text{JSD}_{\text{2-gram}}$) capturing dependencies across consecutive packet pairs, and ($d$) Markov transition matrices ($\text{JSD}_{\text{Markov}}$) capturing first-order transition probabilities between consecutive signed \gls{pl} values.

Furthermore, we assess diversity alignment and exact-sequence overlap between real and synthetic biflow representations through two additional metrics:
\begin{enumerate}[label=($\alph*$), start=5]
    \item \emph{UniqAlign} measures the absolute difference between the uniqueness scores of real and synthetic datasets, where each uniqueness score is defined as the proportion of distinct biflow representations, as proposed in~\cite{aceto2024synthetic}:
    \begin{equation}
    \mathrm{UniqAlign} =
    \left|
    \frac{N_{\mathrm{unique}}^{\mathrm{re}}}{N_{\mathrm{re}}}
    -
    \frac{N_{\mathrm{unique}}^{\mathrm{syn}}}{N_{\mathrm{syn}}}
    \right|,
    \end{equation}
    where $N_{\mathrm{unique}}$ and $N$ denote the number of distinct biflow representations and the total number of biflows, respectively. Lower values indicate better alignment of the relative sequence diversity.
    
    \item \emph{Leakage} measures the exact-sequence overlap between real and synthetic datasets using the overlap score proposed in~\cite{aceto2024synthetic}, based on Jaccard similarity:
    \begin{equation}
        \mathrm{Leakage} =
        \frac{
            |\mathcal{U}_{\mathrm{re}}
            \cap \mathcal{U}_{\mathrm{syn}}|
        }{
            |\mathcal{U}_{\mathrm{re}}
            \cup \mathcal{U}_{\mathrm{syn}}|
        },
    \end{equation}
    where $\mathcal{U}_{\mathrm{re}}$ and $\mathcal{U}_{\mathrm{syn}}$ denote the sets of distinct real and synthetic biflow representations, respectively.
    Lower values indicate less exact-sequence overlap, suggesting reduced memorization of real traffic.
\end{enumerate}

Figure~\ref{fig:radar_fidelity} summarizes the performance of generative models. 
Since all metrics follow a ``lower-is-better'' logic, radar plot axes are inverted, meaning that larger areas indicate better overall scores across the considered metrics.
Results are consistent across \MIRAGE and \TLSETY.

\LLAMA and \MAMBA outperform all alternatives, achieving near-zero \gls{jsd} scores across all evaluated traffic properties.
They effectively capture both marginal \PLDIR distributions ($1$-gram) and more complex sequential dependencies ($2$-gram and Markov).
Notably, \LLAMA achieves the best $\text{JSD}_{\text{Markov}}$, indicating its ability to accurately model first-order transition patterns.
Also, both sequence-based models exhibit near-optimal \emph{UniqAlign} and \emph{Leakage} scores.
These findings indicate that their generated samples closely match real traffic distributions while preserving similar sequence diversity and exhibiting limited exact-sequence overlap (with \MAMBA showing a slight edge over \LLAMA in \emph{Leakage} mitigation).

Interestingly, the \CVAE baseline suitably performs only on coarse-grained properties, 
successfully matching the observed biflow packet-count distribution 
($\text{JSD}_{\text{NumPackets}}\!\approx\!0$). 
However, its fidelity drops when capturing complex patterns, exposing its limitations in modeling fine-grained traffic dynamics, though it still exhibits low \emph{Leakage}.

\NETDIFFUSNR consistently yields the lowest fidelity. 
It struggles to capture structural traffic properties, exhibiting the highest \gls{jsd} scores (\ie the smallest polygon area) and failing to accurately model even the biflow packet counts. 
Despite its poor fidelity, \NETDIFFUSNR exhibits low \emph{Leakage} and moderate \emph{UniqAlign}, highlighting that limited exact-sequence overlap alone does not guarantee realistic traffic generation.

\newcommand{\std}[1]{\text{\tiny$\pm$#1}}

\newcommand{\orangeRF}{\cellcolor[HTML]{F3B562}} %
\newcommand{\orangeCNN}{\cellcolor[HTML]{FCE6C5}} %
\newcommand{\blueRF}{\cellcolor[HTML]{6FC5F3}} %
\newcommand{\blueCNN}{\cellcolor[HTML]{D8F1FC}} %

\begin{table}[ht]
\centering
\caption{
F1-score (\%) of \RF and \CNN  classifiers trained on synthetic traffic and evaluated on real traffic.
Orange and azure denote \MIRAGE and \TLSETY, respectively.
}
\label{tab:rf_f1_comparison_tstr}
\renewcommand{\arraystretch}{1.08}

\resizebox{\linewidth}{!}{
\begin{threeparttable}

\begin{tabular}{lcccc}
\toprule
\multirow{2}{*}{\textbf{\gls{genai} Approach}}
&
\multicolumn{2}{c}{\textbf{\MIRAGE}}
&
\multicolumn{2}{c}{\textbf{\TLSETY}}\\
\cmidrule(lr){2-3}\cmidrule(l){4-5}
&
\orangeRF\textbf{\RF}
&
\orangeCNN\textbf{\CNN}
&
\blueRF\textbf{\RF}
&
\blueCNN\textbf{\CNN}\\
\midrule

\CVAE
&
\orangeRF $66.90\std{0.35}$
&
\orangeCNN $45.17\std{1.73}$
&
\blueRF $75.27\std{0.19}$
&
\blueCNN $63.58\std{0.24}$\\
\NETDIFFUSNR
&
\orangeRF $46.57\std{0.60}$
&
\orangeCNN $43.35\std{1.68}$
&
\blueRF $61.68\std{0.23}$
&
\blueCNN $63.36\std{0.37}$\\
\MAMBA
&
\orangeRF $76.80\std{0.14}$
&
\orangeCNN $60.60\std{1.38}$
&
\blueRF $85.77\std{0.16}$
&
\blueCNN $76.45\std{0.32}$\\
\LLAMA
&
\orangeRF $\textbf{78.83}\std{0.19}$
&
\orangeCNN $\textbf{62.70}\std{1.73}$
&
\blueRF $\textbf{87.18}\std{0.21}$
&
\blueCNN $\textbf{78.42}\std{0.41}$\\

\midrule

\MAMBA w/ rect
&
\orangeRF $77.14\std{0.17}$
&
\orangeCNN $66.36\std{0.42}$
&
\blueRF $85.75\std{0.10}$
&
\blueCNN $78.44\std{0.29}$\\
\LLAMA w/ rect
&
\orangeRF $\textbf{79.58}\std{0.16}$
&
\orangeCNN $\textbf{68.20}\std{0.45}$
&
\blueRF $\textbf{87.90}\std{0.18}$
&
\blueCNN $\textbf{80.37}\std{0.28}$\\

\midrule
Train on Real
&
\orangeRF $85.67$
&
\orangeCNN $66.57$
&
\blueRF $91.77$
&
\blueCNN $84.31$\\
\bottomrule
\end{tabular}

\begin{tablenotes}[flushleft]
\footnotesize
\textbf{Notes:} the synthetic training dataset contains the same number of samples per class as the real dataset, while for \say{w/ rect} all classes have the same number of samples as the majority class (\ie class imbalance rectification); 
results are reported as mean$\pm$standard deviation over $10$ generation runs; 
``Train on Real'' row denotes the real-data reference performance, while boldface identifies the best \gls{genai} model for each dataset--classifier pair.
\end{tablenotes}
\end{threeparttable}
}

\end{table}

\subsection{RA2 -- \gls{ntc}: Train on Synthetic Traffic}
\label{subsec:ra2_tstr}
To address \gls{rq}2, we evaluate generated data utility via a \emph{train-on-synthetic, test-on-real} approach.
We train Random Forest (\RF) and 2D Convolutional Neural Network (\CNN) downstream classifiers exclusively on synthetic samples and evaluate their generalization on unseen real traffic from \MIRAGE and \TLSETY. 
Performance under this setting assesses the ability of synthetic traffic to preserve class-discriminative characteristics, supporting downstream \gls{ntc} without requiring access to the original real training samples.
\gls{ntc} performance is measured using the macro-averaged F1-score, assigning equal weight to each traffic class.

Table~\ref{tab:rf_f1_comparison_tstr} shows that both \RF{} and \CNN{} trained on synthetic traffic (without
class-imbalance rectification) exhibit an expected performance gap relative to real-data training, reflecting the inherent difficulty of fully reproducing realistic traffic characteristics. 
Nevertheless, \LLAMA{}- and \MAMBA{}-generated traffic consistently yields higher classification performance than the \CVAE baseline and \NETDIFFUSNR across both classifiers, including \CNN{}, whose real-data reference performance is lower than that of \RF{}. 
On \MIRAGE, the \RF trained on \LLAMA and \MAMBA samples achieves $78.83\%$ and $76.80\%$ F1-scores, respectively, substantially reducing the gap with the real-data reference ($85.67\%$).

Similarly, on \TLSETY, the synthetic-trained \RF reaches $87.18\%$ (\LLAMA) and $85.77\%$ (\MAMBA) F1-scores, closely trailing the $91.77\%$ obtained with real traffic. 
In contrast, \NETDIFFUSNR severely underperforms sequence-based models across both datasets, confirming that this \gls{dm} struggles to capture traffic characteristics relevant for downstream classification.

\gls{ntc} performance with \LLAMA and \MAMBA can be further improved by leveraging the models' ability to generate an arbitrary number of samples per class, thus enabling \emph{class imbalance rectification}. 
For example, through this class-balancing generation, \LLAMA further improves the \CNN F1-score by $5.5$ percentage points on \MIRAGE and $1.95$ percentage points on \TLSETY.

To summarize, \LLAMA and \MAMBA offer the best balance of generation fidelity and utility for downstream tasks, supporting synthetic-to-real generalization and \gls{ntc} with reduced raw-data sharing while retaining competitive classification performance, whereas \NETDIFFUSNR appears less suitable for realistic traffic generation.

\definecolor{myorange}{HTML}{ff7f0e}
\definecolor{mygreen}{HTML}{2ca02c}
\definecolor{myred}{HTML}{d62728}
\definecolor{myviolet}{HTML}{9467bd}

\begin{figure}[t]
\begin{minipage}{\columnwidth}
    \centering
    \includegraphics[width=.8\linewidth]{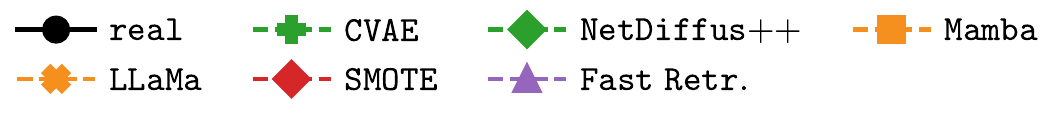}

    \subfloat[\MIRAGE]{%
      \begin{tabular}{@{}c@{}}
        \includegraphics[height=0.325\linewidth,trim=10 10 180 30,clip]{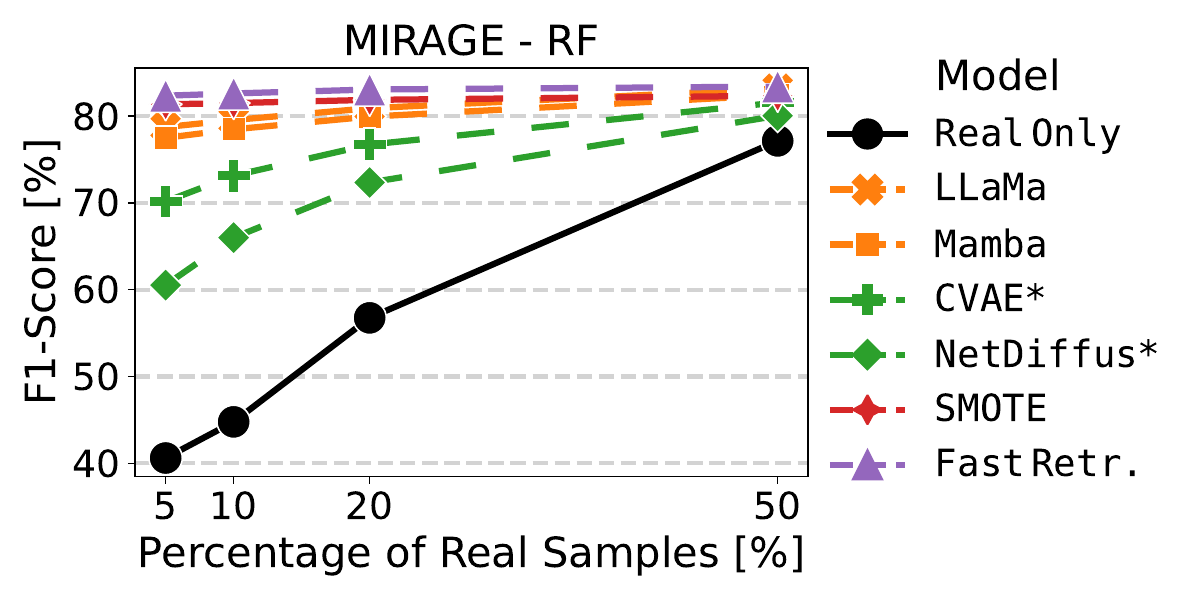}\\[2pt]
        \includegraphics[height=0.350\linewidth,trim=10 10 180 10,clip]{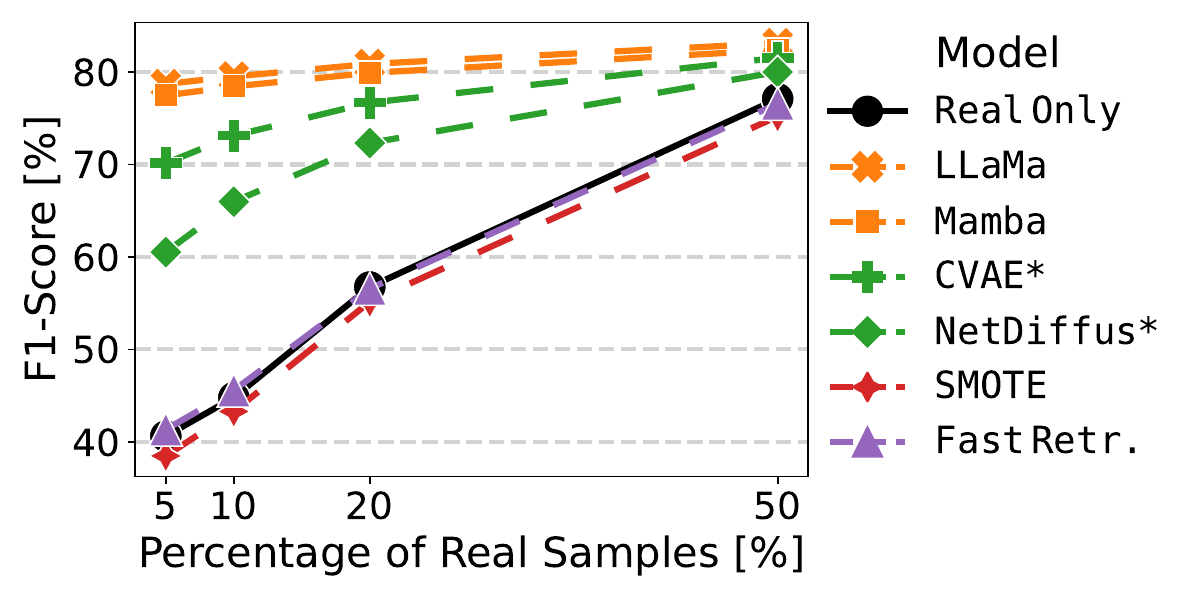}\\[2pt]
        \includegraphics[height=0.405\linewidth]{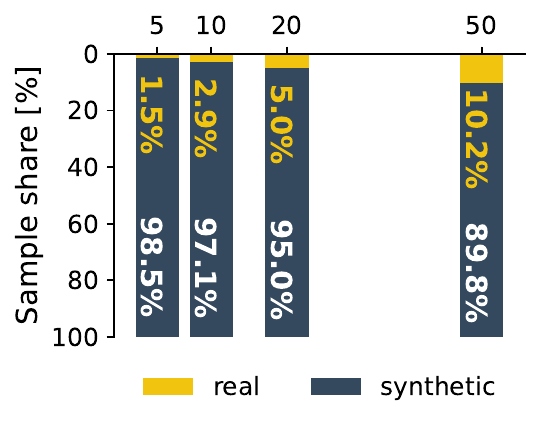}
      \end{tabular}}
    \hfil
    \subfloat[\TLSETY]{%
      \begin{tabular}{@{}c@{}}
        \includegraphics[height=0.325\linewidth,trim=30 10 180 30,clip]{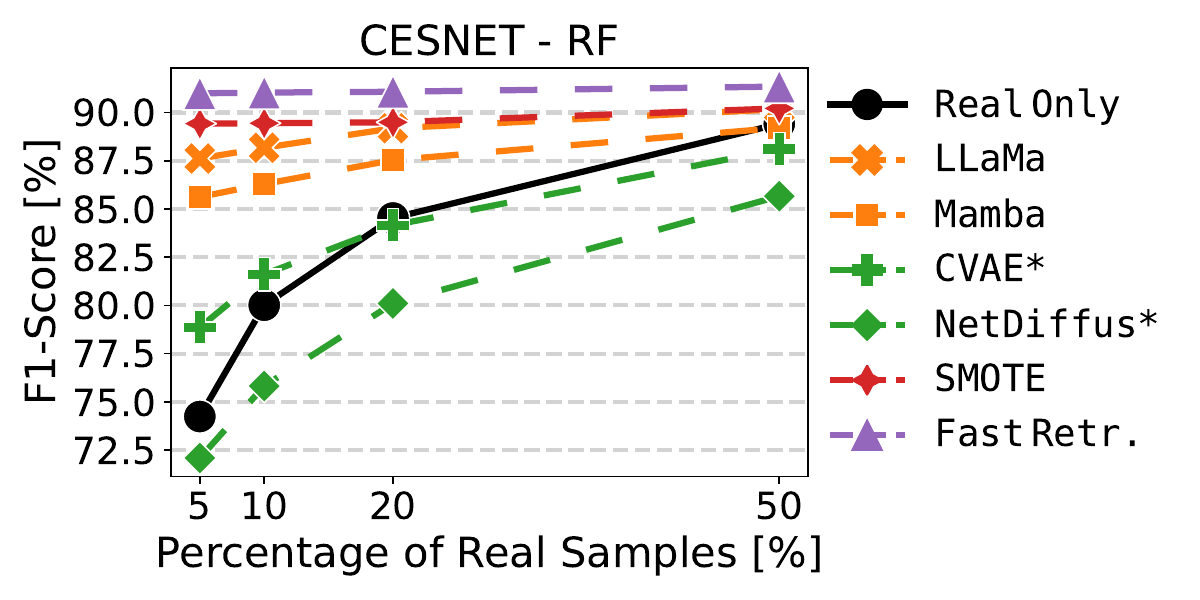}\\[2pt]
        \includegraphics[height=0.350\linewidth,trim=30 10 180 10,clip]{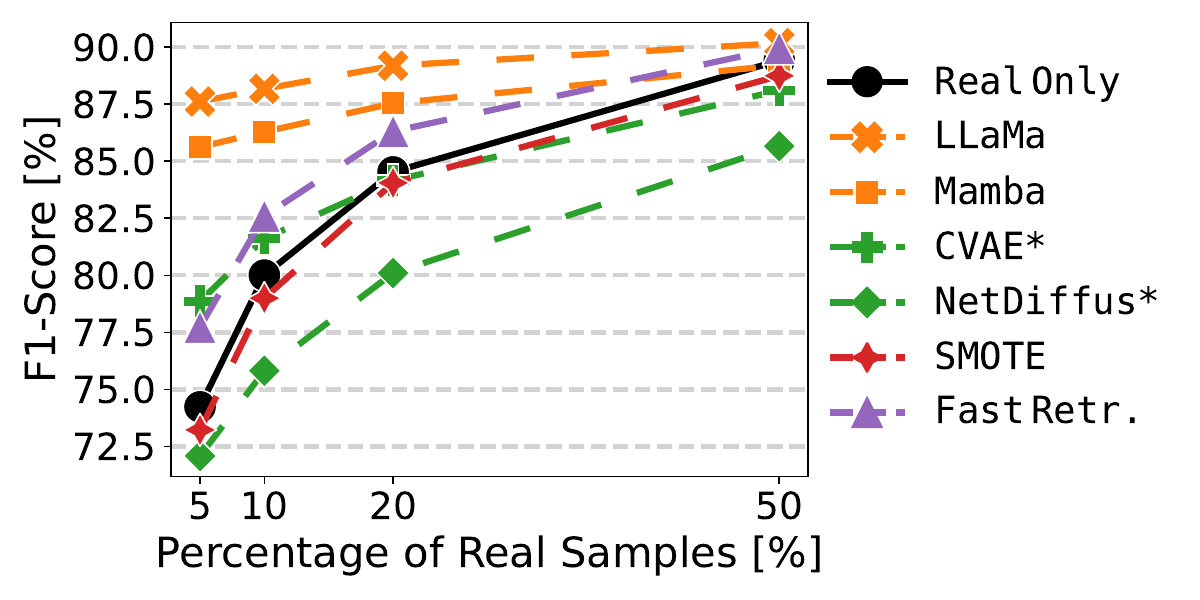}\\[2pt]
        \includegraphics[height=0.405\linewidth, trim=22 0 0 0, clip]{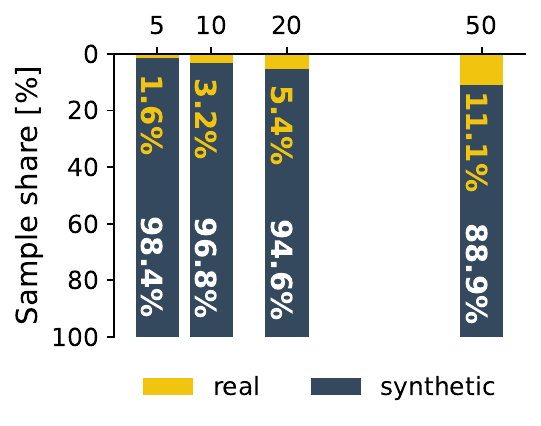}
      \end{tabular}}
    
    \caption{
    Data augmentation under low-data regimes for \MIRAGE{} (left)
    and \TLSETY{} (right).
    The top and middle panels report the F1-score of an \RF{}
    classifier when conventional augmentation techniques have
    access to the full and limited real training data,
    respectively, at augmentation time, while \gls{genai}
    models are pretrained on the full training set in both
    settings.
    The bottom panels show the real/synthetic composition
    of the augmented training set, shared by both settings.
    The x-axis denotes the percentage of real training
    samples retained by uniformly downsampling each class;
    synthetic samples are then added to upsample each class
    to the size of the original majority class, yielding
    a balanced training set.
    Colors in the top and middle panels indicate the
    approach family:
    \textbf{\textcolor{myorange}{orange}} for sequence-based
    models,
    \textbf{\textcolor{mygreen}{green}} for diffusion and
    variational models,
    \textbf{\textcolor{myred}{red}} for statistical techniques,
    \textbf{\textcolor{myviolet}{violet}} for expert
    transformations, and
    \textbf{black} for real-only training.
    }
    \label{fig:f1-exec}
\end{minipage}
\end{figure}

\subsection{RA3 -- \gls{ntc}: Data Augmentation}
\label{subsec:ra3_augmentation}
To answer \gls{rq}3, we investigate \gls{genai}-driven data augmentation under data-scarcity conditions, where only a limited fraction of real training samples is available to the downstream classifier. 
Specifically, we uniformly retain $5\%$, $10\%$, $20\%$, or $50\%$ of the real training samples from each class and augment the resulting subsets with synthetic samples to balance all classes up to the size of the original majority class.
This setting is particularly relevant given the class imbalance observed in both datasets (see Fig.~\ref{fig:dataset}).
\gls{genai} models are trained offline on the full labeled training set and subsequently used to generate class-conditioned synthetic biflows without requiring access to the original training traces at augmentation time. 
This reflects the practical scenario of a network operator leveraging a pre-trained \gls{genai} model to compensate for limited labeled data.

The downstream classifier (\RF) is trained on the retained real samples together with the generated traffic.
We compare \gls{genai} models against two conventional (non-\gls{genai}) augmentation baselines: 
($i$) \emph{Fast Retransmit}~\cite{wang2024data}, a \emph{domain-expert traffic transformation}, 
and ($ii$) \emph{SMOTE}, a \emph{statistical oversampling} technique.
In more detail, Fast Retransmit probabilistically delays a single packet to mimic a TCP retransmission, while SMOTE interpolates existing samples in the feature space.
Unlike generative models, which learn traffic distributions during offline training and can subsequently synthesize samples independently, \emph{Fast Retransmit and SMOTE require access to real reference samples at augmentation time}.

To highlight the implications of this distinction, we consider \emph{two experimental settings} differing in the amount of real reference data available to conventional augmentation techniques, as illustrated in Fig.~\ref{fig:f1-exec}.
In the first setting (top panels), conventional augmentation techniques have access to the full real training dataset as reference data for synthetic sample generation, while \gls{genai} models exploit the distributions learned from the same dataset during offline training.
This setting provides a comparison in which all methods can leverage information from the entire training set.
In the second setting (middle panels), only the reduced real training subsets are available at augmentation time.
While SMOTE and Fast Retransmit must generate synthetic samples from these limited reference data, pretrained \gls{genai} models can still sample from the distributions learned offline, without requiring access to the original training traces.
This setting highlights an additional operational capability of pretrained \gls{genai} models: enabling traffic augmentation even when the full original training dataset is no longer available. 
In both settings, the downstream classifier receives the same proportions of real and synthetic samples at each retention level, as illustrated in the bottom panels of Fig.~\ref{fig:f1-exec}. 
As the fraction of retained real data decreases, synthetic traffic progressively dominates the balanced training set, accounting for more than $95\%$ of the samples in the most challenging low-data regimes.

When conventional augmentation methods have access to the full training set (top panels), both SMOTE and Fast Retransmit benefit from the availability of abundant real reference samples.
In particular, Fast Retransmit achieves F1-scores comparable to, or even higher than, those obtained with sequence-based \gls{genai} models in several configurations.
These results show that conventional augmentation techniques can be highly competitive when the original training data remain accessible.

However, a different picture emerges when only limited real reference data are available at augmentation time (middle panels).
In this setting, pretrained \gls{genai} models retain the ability to generate synthetic traffic from distributions learned offline, whereas conventional techniques must rely on the reduced set of available real samples.
This operational advantage translates into larger classification gains for sequence-based \gls{genai} models, particularly in the most challenging low-data regimes, where only $5$--$20\%$ of the original training samples are retained.
On \MIRAGE, \LLAMA- and \MAMBA-generated samples substantially improve classification performance over real-only training, rapidly approaching the F1-scores achieved with abundant real data.
On \TLSETY, the gains are more moderate but remain consistent, with \LLAMA increasing the F1-score from approximately $74\%$ to $88\%$ at the $5\%$ retention level. 
In contrast, SMOTE provides limited benefits, while Fast Retransmit yields more noticeable gains on \TLSETY.
Among the other generative architectures, \CVAE generally improves upon real-only training, whereas \NETDIFFUSNR shows less consistent results: it improves performance on \MIRAGE, although remaining well below sequence-based models, and systematically underperforms on \TLSETY.

Overall, these findings highlight a distinctive operational advantage of pretrained \gls{genai} models: they enable effective traffic augmentation by reusing previously learned distributions even when the full original training set is unavailable.

\definecolor{TrainCol}{RGB}{230,236,255}   %
\definecolor{GenCol}{RGB}{232,249,232}     %
\definecolor{SizeCol}{RGB}{248,233,242}    %
\definecolor{RowGray}{RGB}{250,250,250}
\definecolor{LlamaRow}{RGB}{255,248,230}    %

\begin{table}[t]
\centering
\caption{
Training and generation resource requirements (time, utilization, memory, and on-disk size) of the considered \gls{genai} models. Quantized \texttt{LLaMA} variants are included to assess deployment-oriented optimization.}
\label{tab:resource-usage}

\setlength{\tabcolsep}{4pt}

\resizebox{\linewidth}{!}{
\begin{threeparttable}

\rowcolors{3}{white}{RowGray}

\begin{tabular}{
l|
>{\columncolor{TrainCol}}c
>{\columncolor{TrainCol}}c
>{\columncolor{TrainCol}}c|
>{\columncolor{GenCol}}c
>{\columncolor{GenCol}}c
>{\columncolor{GenCol}}c|
>{\columncolor{SizeCol}}c
}

\toprule
\multirow{2}{*}{\textbf{Model}} &
\multicolumn{3}{c|}{\cellcolor{TrainCol}\textbf{Training (GPU)}} &
\multicolumn{3}{c|}{\cellcolor{GenCol}\textbf{Generation (RPi 4B)}} &
\multicolumn{1}{c}{\cellcolor{SizeCol}\textbf{Model Size}} \\

&
\makecell[c]{\textbf{Time}\\$[$s/epoch$]$} &
\makecell[c]{\textbf{GPU}\\$[\%]$} &
\makecell[c]{\textbf{Mem}\\$[$MB$]$} &
\makecell[c]{\textbf{Time}\\$[$s/sample$]$} &
\makecell[c]{\textbf{CPU}\\$[\%]$} &
\makecell[c]{\textbf{Mem}\\$[$MB$]$} &
\makecell[c]{\textbf{\, $\phantom{0}$on Disk$\phantom{0}$\qquad}\\$[$MB$]$} \\

\midrule

$\mathtt{CVAE}$ &
22.57 & 18 & 379 &
0.07 & 30 & 402 &
3.9 \\

\NETDIFFUSNR &
117.03 & 21 & 387 &
105.67 & 61 & 339 &
4.2 \\

\MAMBA &
108.48 & 15 & 377 &
10.18 & 47 & 534 &
15.5 \\

\rowcolor{LlamaRow}
\LLAMA &
36.81 & 20 & 393 &
2.83 & 44 & 499 &
7.9 \\

\midrule

\rowcolor{LlamaRow}
\LLAMA$\!_{\mathtt{int8}\text{-}\mathtt{WO}}$ &
\multicolumn{3}{c|}{same as \LLAMA} &
3.48 & 45 & 509 &
3.5 \\

\rowcolor{LlamaRow}
\LLAMA$\!_{\mathtt{int8}\text{-}\mathtt{DA}}$ &
\multicolumn{3}{c|}{same as \LLAMA} &
2.99 & 41 & 497 &
3.4 \\

\bottomrule

\end{tabular}

\begin{tablenotes}[flushleft]
\textbf{Training:}
measured on a high-end datacenter GPU (48\,GB memory);
hyperparameters: 5 classes, 500 samples/class, 10 epochs, batch size of 1.

\textbf{Generation:}
measured on an emulated Raspberry Pi 4 Model B (CPU-only, 4\,GB memory);
hyperparameters: 5 classes, 10 samples/class (50 samples total), batch size of 1.

\textbf{Notes:}
all values are the median over 10 runs;
results for larger datasets scale approximately with the number of classes and samples.
\end{tablenotes}
\end{threeparttable}
}

\end{table}

\subsection{RA4 -- Computational Efficiency}
\label{subsec:ra4_computational}
To address \gls{rq}4, Tab.~\ref{tab:resource-usage} evaluates lightweight \gls{genai} models in terms of training time, generation latency, GPU/CPU utilization, memory usage, and on-disk footprint.
Models are trained on a high-end datacenter GPU and evaluated for inference on an emulated Raspberry Pi 4B (RPi~4B) to assess deployment-oriented generation under constrained, CPU-only conditions.%
\footnote{We emulate the RPi~4B via a QEMU VM with 64-bit ARMv8-A, four Cortex-A72 cores, $4\,$GB RAM, and Ubuntu~22.04.5~LTS.}
The results reveal substantial differences in computational costs across architectures, highlighting the importance of balancing generation fidelity and resource requirements for practical deployment.

Training requires between $22.57$ and $117.03$\,s/epoch across models, with \LLAMA requiring only $36.8$\,s/epoch despite offering the highest generation fidelity. 
During training, memory consumption remains below $400$\,MB across all models. 
GPU utilization is similarly modest, ranging from $15\%$ (\MAMBA) to $21\%$ (\NETDIFFUSNR), reflecting limited accelerator demands in this setting.
Architectural differences become even clearer during CPU-only inference. 
Among the recent \gls{genai} alternatives, \LLAMA achieves the lowest latency ($2.83$\,s/sample), compared with $10.18$\,s/sample for \MAMBA and $105.67$\,s/sample for \NETDIFFUSNR. 
It also exhibits slightly lower CPU utilization ($44\%$ vs.~$47\%$) and memory consumption ($499$ vs.~$534$\,MB) than \MAMBA, further strengthening its suitability for deployment.
Lastly, all models retain a compact storage footprint ($3.9-15.5$\,MB), supporting deployment on resource-constrained edge devices.

These findings identify \LLAMA as the strongest candidate for deployment, motivating a \gls{ptq} analysis. 
We evaluate two \texttt{int8} variants: \emph{weight-only} (\LLAMA$\!_{\mathtt{int8}\text{-}\mathtt{WO}}$), which quantizes only weights, and \emph{dynamic activation} (\LLAMA$\!_{\mathtt{int8}\text{-}\mathtt{DA}}$), which also quantizes activations during inference.
Both preserve fidelity and downstream classification performance while reducing model size by more than $2\times$, from $7.9$\,MB to $3.5$\,MB and $3.4$\,MB, respectively.
Since the compression is nearly identical, the main difference lies in inference latency: \LLAMA$\!_{\mathtt{int8}\text{-}\mathtt{DA}}$ achieves lower generation latency ($2.99$\,s/sample) than \LLAMA$\!_{\mathtt{int8}\text{-}\mathtt{WO}}$ ($3.48$\,s/sample).
Therefore, lightweight \gls{genai} architectures, particularly quantized \LLAMA, combine high-quality \gls{ntg} with a compact on-disk footprint and practical computational costs. 
Overall, these results demonstrate that high-fidelity traffic generators can be designed with modest resource requirements, making them suitable for practical deployment.

\section{Conclusion}
\label{sec:conclusion}
This work investigated lightweight \gls{genai} models for \gls{ntg}, focusing on traffic fidelity, synthetic-only training, data augmentation in low-data regimes, and deployment feasibility.
Our results show that sequence-based models, particularly lightweight \LLAMA and \MAMBA, are the most promising solutions for realistic traffic generation, classification with reduced raw-data sharing, and effective data augmentation.
In fidelity assessment (\gls{rq}1), \LLAMA and \MAMBA achieve near-zero \gls{jsd} scores across all traffic properties.
For downstream \gls{ntc}, \LLAMA enables synthetic-only training with an absolute F1-score gap below $7\%$ compared with real-data training (\gls{rq}2), and improves low-data classification by up to $40\%$ through augmentation, even without continued access to the full original training set (\gls{rq}3).
Regarding deployment feasibility (\gls{rq}4), \LLAMA offers the best fidelity--efficiency trade-off, combining sub-$40$\,s/epoch training and $2.83$\,s/sample generation latency, while post-training quantization reduces its on-disk model size to as little as $3.4$\,MB.
Conversely, the diffusion-based \NETDIFFUSNR and the \CVAE baseline proved less effective, either incurring prohibitive generation latencies or failing to capture fine-grained packet-sequence dependencies.

\subsec{Future Directions.}
The present study focuses on lightweight \gls{genai} models trained on compact flow-level traffic representations and evaluated on two public benchmark datasets.
Future work will investigate broader traffic domains, adaptive generation strategies, and hybrid pipelines combining generative models with domain-specific traffic transformations.
We also plan to evaluate lightweight \gls{genai} architectures in real-world deployment scenarios, including the integration of quantized models into edge-based traffic analysis systems.
Finally, we will explore lightweight agentic-AI frameworks to enable autonomous network monitoring, closed-loop network management, and proactive cyber-defense.

\footnotesize

\bibliographystyle{unsrtnat}
\bibliography{IEEEabrv,tc_bibliography}

\begin{thebibliography}{26}
\providecommand{\natexlab}[1]{#1}
\providecommand{\url}[1]{\texttt{#1}}
\expandafter\ifx\csname urlstyle\endcsname\relax
  \providecommand{\doi}[1]{doi: #1}\else
  \providecommand{\doi}{doi: \begingroup \urlstyle{rm}\Url}\fi

\bibitem[Bariah et~al.(2024)Bariah, Zhao, Zou, Tian, Bader, and Debbah]{bariah2024large}
Lina Bariah, Qiyang Zhao, Hang Zou, Yu~Tian, Faouzi Bader, and Merouane Debbah.
\newblock {Large generative {AI} models for telecom: The next big thing?}
\newblock \emph{{IEEE Communications Magazine}}, 62\penalty0 (11):\penalty0 84--90, 2024.
\newblock \doi{10.1109/MCOM.001.2300364}.

\bibitem[Xu et~al.(2025)Xu, Mu, Liu, Xing, Liu, and Nallanathan]{xu2025generative}
Xiaoxia Xu, Xidong Mu, Yuanwei Liu, Hong Xing, Yue Liu, and Arumugam Nallanathan.
\newblock {Generative Artificial Intelligence for Mobile Communications: A Diffusion Model Perspective}.
\newblock \emph{{IEEE Communications Magazine}}, 63\penalty0 (7):\penalty0 98--105, 2025.
\newblock \doi{10.1109/MCOM.001.2400284}.

\bibitem[Long et~al.(2025)Long, Tang, Li, Tan, Jin, Zhao, and Kato]{long20246g}
Sifan Long, Fengxiao Tang, Yangfan Li, Tiao Tan, Zhengjie Jin, Ming Zhao, and Nei Kato.
\newblock {{6G} Comprehensive Intelligence: Network Operations and Optimization Based on Large Language Models}.
\newblock \emph{{IEEE Network}}, 39\penalty0 (4):\penalty0 192--201, 2025.
\newblock \doi{10.1109/MNET.2024.3470774}.

\bibitem[Bovenzi et~al.(2025)Bovenzi, Cerasuolo, Ciuonzo, Di~Monda, Guarino, Montieri, Persico, and Pescap{\'e}]{bovenzi2025mapping}
Giampaolo Bovenzi, Francesco Cerasuolo, Domenico Ciuonzo, Davide Di~Monda, Idio Guarino, Antonio Montieri, Valerio Persico, and Antonio Pescap{\'e}.
\newblock {Mapping the landscape of generative {AI} in network monitoring and management}.
\newblock \emph{{IEEE Transactions on Network and Service Management}}, 22\penalty0 (3):\penalty0 2441--2472, 2025.
\newblock \doi{10.1109/TNSM.2025.3543022}.

\bibitem[P{\'e}rez-Jove et~al.(2026)P{\'e}rez-Jove, Munteanu, Dorado, Pazos, and V{\'a}zquez-Naya]{perez2026network}
Rub{\'e}n P{\'e}rez-Jove, Cristian~R Munteanu, Juli{\'a}n Dorado, Alejandro Pazos, and Jose V{\'a}zquez-Naya.
\newblock {Network traffic foundation models: A systematic review}.
\newblock \emph{{Computer Networks}}, page 111998, 2026.
\newblock \doi{10.1016/j.comnet.2026.111998}.

\bibitem[Aceto et~al.(2024{\natexlab{a}})Aceto, Ciuonzo, Montieri, Persico, and Pescap{\'e}]{aceto2023ai}
Giuseppe Aceto, Domenico Ciuonzo, Antonio Montieri, Valerio Persico, and Antonio Pescap{\'e}.
\newblock {AI-Powered Internet Traffic Classification: Past, Present, and Future}.
\newblock \emph{{IEEE Communications Magazine}}, 62\penalty0 (9):\penalty0 168--175, 2024{\natexlab{a}}.
\newblock \doi{10.1109/MCOM.001.2300361}.

\bibitem[Aceto et~al.(2021)Aceto, Bovenzi, Ciuonzo, Montieri, Persico, and Pescap{\'e}]{aceto2021characterization}
Giuseppe Aceto, Giampaolo Bovenzi, Domenico Ciuonzo, Antonio Montieri, Valerio Persico, and Antonio Pescap{\'e}.
\newblock {Characterization and prediction of mobile-app traffic using {Markov} modeling}.
\newblock \emph{{IEEE Transactions on Network and Service Management}}, 18\penalty0 (1):\penalty0 907--925, 2021.
\newblock \doi{10.1109/TNSM.2021.3051381}.

\bibitem[Aceto et~al.(2024{\natexlab{b}})Aceto, Giampaolo, Guida, Izzo, Pescap{\`e}, Piccialli, and Prezioso]{aceto2024synthetic}
Giuseppe Aceto, Fabio Giampaolo, Ciro Guida, Stefano Izzo, Antonio Pescap{\`e}, Francesco Piccialli, and Edoardo Prezioso.
\newblock {Synthetic and Privacy-Preserving Traffic Trace Generation using Generative {AI} Models for Training Network Intrusion Detection Systems}.
\newblock \emph{{Journal of Network and Computer Applications}}, page 103926, 2024{\natexlab{b}}.
\newblock \doi{10.1016/j.jnca.2024.103926}.

\bibitem[Meslet-Millet et~al.(2022)Meslet-Millet, Mouysset, and Chaput]{meslet2022necstgen}
Fabien Meslet-Millet, Sandrine Mouysset, and Emmanuel Chaput.
\newblock {NeCSTGen: An approach for realistic network traffic generation using Deep Learning}.
\newblock In \emph{GLOBECOM 2022 - 2022 IEEE Global Communications Conference}, pages 3108--3113, 2022.
\newblock \doi{10.1109/GLOBECOM48099.2022.10000731}.

\bibitem[Gudovskiy et~al.(2022)Gudovskiy, Ishizaka, and Kozuka]{gudovskiy2022cflow}
Denis Gudovskiy, Shun Ishizaka, and Kazuki Kozuka.
\newblock {{CFLOW-AD}: Real-time unsupervised anomaly detection with localization via conditional normalizing flows}.
\newblock In \emph{IEEE/CVF Winter Conf. on Applications of Computer Vision (WACV)}, pages 98--107, 2022.
\newblock \doi{10.1109/WACV51458.2022.00188}.

\bibitem[Madarasingha et~al.(2022)Madarasingha, Muramudalige, Jourjon, Jayasumana, and Thilakarathna]{madarasingha2022videotrain}
Chamara Madarasingha, Shashika~R Muramudalige, Guillaume Jourjon, Anura Jayasumana, and Kanchana Thilakarathna.
\newblock {VideoTrain++: GAN-based adaptive framework for synthetic video traffic generation}.
\newblock \emph{{Computer Networks}}, 206:\penalty0 108785, 2022.
\newblock \doi{10.1016/j.comnet.2022.108785}.

\bibitem[Hui et~al.(2022)Hui, Wang, Wang, Yang, Liu, Jin, and Li]{hui2022knowledge}
Shuodi Hui, Huandong Wang, Zhenhua Wang, Xinghao Yang, Zhongjin Liu, Depeng Jin, and Yong Li.
\newblock {Knowledge enhanced {GAN} for {IoT} traffic generation}.
\newblock In \emph{ACM Web Conference (WWW)}, pages 3336--3346, 2022.
\newblock \doi{10.1145/3485447.3511976}.

\bibitem[Bikmukhamedov and Nadeev(2021)]{bikmukhamedov2021}
R.~F. Bikmukhamedov and A.~F. Nadeev.
\newblock {Multi-Class Network Traffic Generators and Classifiers Based on Neural Networks}.
\newblock In \emph{Systems of Signals Generating and Proc. in the Field of on Board Comm.}, 2021.
\newblock \doi{10.1109/IEEECONF51389.2021.9416067}.

\bibitem[Kholgh and Kostakos(2023)]{Kholgh2023PACGPT}
Danial~Khosh Kholgh and Panos Kostakos.
\newblock {{PAC-GPT}: A Novel Approach to Generating Synthetic Network Traffic With {GPT-3}}.
\newblock \emph{{IEEE Access}}, 11:\penalty0 114936--114951, 2023.
\newblock \doi{10.1109/ACCESS.2023.3325727}.

\bibitem[Qu et~al.(2024)Qu, Ma, and Li]{qu2024trafficgpt}
Jian Qu, Xiaobo Ma, and Jianfeng Li.
\newblock {{TrafficGPT}: Breaking the Token Barrier for Efficient Long Traffic Analysis and Generation}.
\newblock \emph{{arXiv preprint arXiv:2403.05822}}, 2024.

\bibitem[Cui et~al.(2025)Cui, Lin, Li, Chen, Yin, Li, and Xu]{cui2025trafficllm}
Tianyu Cui, Xinjie Lin, Sijia Li, Miao Chen, Qilei Yin, Qi~Li, and Ke~Xu.
\newblock {{TrafficLLM}: Enhancing large language models for network traffic analysis with generic traffic representation}.
\newblock \emph{{arXiv preprint arXiv:2504.04222}}, 2025.

\bibitem[Mayhoub et~al.(2026)Mayhoub, Foh, Mashhadi, Shojafar, and Tafazolli]{mayhoub2026talk}
Samara Mayhoub, Chuan~Heng Foh, Mahdi~Boloursaz Mashhadi, Mohammad Shojafar, and Rahim Tafazolli.
\newblock {Talk Like a Packet: Rethinking Network Traffic Analysis with Transformer Foundation Models}.
\newblock \emph{{IEEE Communications Magazine}}, 64\penalty0 (5):\penalty0 36--42, 2026.
\newblock \doi{10.1109/MCOM.001.2500414}.

\bibitem[Delgado-Soto et~al.(2025)Delgado-Soto, de~Vergara, Gonz{\'a}lez, Perdices, and de~Pedro]{delgado2025gpt}
Javier~Aday Delgado-Soto, Jorge E~L{\'o}pez de~Vergara, Iv{\'a}n Gonz{\'a}lez, Daniel Perdices, and Luis de~Pedro.
\newblock Gpt on the wire: Towards realistic network traffic conversations generated with large language models.
\newblock \emph{Computer Networks}, 265:\penalty0 111308, 2025.
\newblock \doi{10.1016/j.comnet.2025.111308}.

\bibitem[Chu et~al.(2024)Chu, Jiang, Liu, Bhagoji, Bronzino, Schmitt, and Feamster]{chu2024feasibility}
Andrew Chu, Xi~Jiang, Shinan Liu, Arjun Bhagoji, Francesco Bronzino, Paul Schmitt, and Nick Feamster.
\newblock {Feasibility of state space models for network traffic generation}.
\newblock In \emph{Proc. of the SIGCOMM Workshop on Networks for AI Computing}, pages 9--17, 2024.
\newblock \doi{10.1145/3672198.3673792}.

\bibitem[Sivaroopan et~al.(2024)Sivaroopan, Bandara, Madarasingha, Jourjon, Jayasumana, and Thilakarathna]{sivaroopan2023netdiffus}
Nirhoshan Sivaroopan, Dumindu Bandara, Chamara Madarasingha, Guillaume Jourjon, Anura~P. Jayasumana, and Kanchana Thilakarathna.
\newblock {{NetDiffus}: Network Traffic Generation by Diffusion Models through Time-Series Imaging}.
\newblock \emph{{Computer Networks}}, 251:\penalty0 110616, 2024.
\newblock ISSN 1389-1286.
\newblock \doi{https://doi.org/10.1016/j.comnet.2024.110616}.

\bibitem[Jiang et~al.(2024)Jiang, Liu, Gember-Jacobson, Bhagoji, Schmitt, Bronzino, and Feamster]{jiang2024netdiffusion}
Xi~Jiang, Shinan Liu, Aaron Gember-Jacobson, Arjun~Nitin Bhagoji, Paul Schmitt, Francesco Bronzino, and Nick Feamster.
\newblock {NetDiffusion: Network Data Augmentation Through Protocol-Constrained Traffic Generation}.
\newblock \emph{{Proc. of the ACM on Measurement and Analysis of Computing Systems}}, 8\penalty0 (1):\penalty0 1--32, 2024.

\bibitem[Wang et~al.(2024)Wang, Finamore, Michiardi, Gallo, and Rossi]{wang2024data}
Chao Wang, Alessandro Finamore, Pietro Michiardi, Massimo Gallo, and Dario Rossi.
\newblock {Data Augmentation for Traffic Classification}.
\newblock In \emph{Int. Conf. on Passive and Active Network Measurement}, pages 159--186. Springer, 2024.
\newblock \doi{10.1007/978-3-031-56249-5_7}.

\bibitem[Touvron et~al.(2023)Touvron, Lavril, Izacard, Martinet, Lachaux, and Lacroix{, et al.}]{touvron2023llama}
Hugo Touvron, Thibaut Lavril, Gautier Izacard, Xavier Martinet, Marie-Anne Lachaux, and Timoth{\'e}e Lacroix{, et al.}
\newblock {LLaMA: Open and Efficient Foundation Language Models}.
\newblock \emph{{arXiv preprint arXiv:2302.13971}}, 2023.

\bibitem[Aceto et~al.(2019)Aceto, Ciuonzo, Montieri, Persico, and Pescap{\'e}]{aceto2019mirage}
Giuseppe Aceto, Domenico Ciuonzo, Antonio Montieri, Valerio Persico, and Antonio Pescap{\'e}.
\newblock {MIRAGE: Mobile-app traffic capture and ground-truth creation}.
\newblock In \emph{2019 4th International Conference on Computing, Communications and Security (ICCCS)}, pages 1--8. IEEE, 2019.
\newblock \doi{10.1109/CCCS.2019.8888137}.

\bibitem[Luxemburk and {\v{C}}ejka(2023)]{luxemburk2023fine}
Jan Luxemburk and Tom{\'a}{\v{s}} {\v{C}}ejka.
\newblock {Fine-grained TLS services classification with reject option}.
\newblock \emph{{Computer Networks}}, 220:\penalty0 109467, 2023.
\newblock \doi{10.1016/j.comnet.2022.109467}.

\bibitem[Nascita et~al.(2021)Nascita, Montieri, Aceto, Ciuonzo, Persico, and Pescap{\'e}]{nascita2021xai}
Alfredo Nascita, Antonio Montieri, Giuseppe Aceto, Domenico Ciuonzo, Valerio Persico, and Antonio Pescap{\'e}.
\newblock {XAI meets mobile traffic classification: Understanding and improving multimodal deep learning architectures}.
\newblock \emph{{IEEE Transactions on Network and Service Management}}, 18\penalty0 (4):\penalty0 4225--4246, 2021.
\newblock \doi{10.1109/TNSM.2021.3098157}.

\end{thebibliography}

\end{document}